\documentclass[prb,nopacs,twocolumn,preprintnumbers,amsmath,amssymb]{revtex4}

\usepackage{graphicx}% Include figure files
\usepackage{dcolumn}% Align table columns on decimal point
\usepackage{color}
\usepackage{bm}% bold math

\begin{document}

\title{Generating quantizing pseudomagnetic fields by bending graphene ribbons}
\author{F. Guinea$^1$, A. K. Geim$^2$,  M. I. Katsnelson$^3$, K. S. Novoselov$^4$}
{\affiliation{$^1$ Instituto de Ciencia de Materiales de Madrid,
CSIC, Sor Juana In\'es de la Cruz 3  E28049 Madrid, Spain \\ $^2$
Centre for Mesoscience and Nanotechnology,
University of Manchester, Manchester M13 9PL, United Kingdom \\
$^3$ Institute for Molecules and Materials, Radboud University of
Nijmegen, Heyendaalseweg 135, NL-6525 AJ, Nijmegen, The Netherlands
\\$^4$ School of Physics \& Astronomy, University of Manchester, Manchester, M13 9PL, United Kingdom
}
\begin{abstract}
We analyze the mechanical deformations that are required to create
uniform pseudomagnetic fields  in graphene. It is shown that, if a
ribbon is bent in-plane into a circular arc, this can lead to
fields exceeding 10T, which is sufficient for the observation of
pseudo-Landau quantization. The arc geometry is simpler than those
suggested previously and, in our opinion, has much better chances
to be realized experimentally soon. The effects of a scalar
potential induced by dilatation in this geometry is shown to be
negligible.
\end{abstract}
%\pacs{73.20.-r; 73.20.Hb; 73.23.-b; 73.43.-f}

\maketitle Graphene exhibits a number of unique features not found
in conventional metals and insulators.\cite{NGPNG09,AGSci09} Among
them is the possibility to stretch graphene elastically  by more
than 15$\%$\cite{Hone08}, and to control in different ways the
induced
strains\cite{Betal08,Betal08b,Ketal09,Tetal09,Metal09,Betal09,Hetal09,Tetal09b}.
This offers a prospect of tuning electronic characteristics of
graphene devices not only by external electric field but also by
mechanical strain\cite{AGSci09,uniax09}, a possibility being
extensively discussed
theoretically\cite{uniax09,AV08,FGK08,PN09,BK09,CJS09,VPP09,MPMT09,Retal09,FR09}.
In particular, the presence of two valleys at the opposite corners
of graphene's Brillouin zone implies that long wavelength lattice
deformations induce an effective gauge field acting on the electrons
and holes, which has the opposite sign for the two
valleys.\cite{SA02b,M07,NGPNG09} This yields an enticing possibility
of creating such gauge fields that would mimic a uniform magnetic
field \emph{B} and, consequently, generate energy gaps in the
electronic spectrum and lead to a zero-\emph{B} analogue of the
quantum Hall effect.\cite{GKG09} Both isotropic and uniaxial strains
result\cite{GKG09,uniax09} in zero pseudomagnetic field $B_{S}$ but
as shown recently\cite{GKG09} deformations with a triangular
symmetry can lead to strong uniform $B_{S}$. Moreover, the
strained-induced pseudomagnetic field can easily reach quantizing
values, exceeding 10T in submicron devices for deformations less
than $10\%$.\cite{GKG09}

Unfortunately, all the  geometries of applied strain, which were
suggested  previously,\cite{GKG09} are rather difficult to realize
experimentally. In this Communication, we report an alternative
strain configuration that does not require a complex triangular
symmetry and, in fact, is a straightforward extension of the
geometry typically used in experimental studies of strained
devices\cite{Metal09,Hetal09,Tetal09b}. We have found that simple
in-plane bending of graphene ribbons (see Fig.~\ref{sketch}) should
lead to strong practically uniform $B_{S}$. We believe that this
finding can speed up the observation of the pseudomagnetic quantum
Hall effect and related phenomena.
\begin{figure}
\includegraphics[width=6cm]{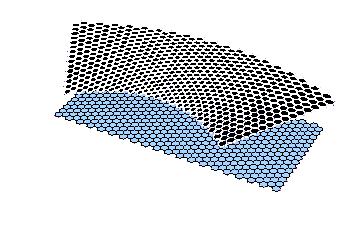}
\caption[fig]{(Color online). Sketch of the suggested bending geometry
that would generate a uniform pseudomagnetic field and open band gaps
in graphene's electronic spectrum.
The graphene rectangle (lower image) is bent into a circular arc (upper).}
 \label{sketch}
\end{figure}

First, let us complete the analysis of Ref.\onlinecite{GKG09} by
classifying the strain distributions that are compatible with
equilibrium elasticity and lead to a uniform pseudomagnetic field.

We will use the coordinates that are fixed with respect  to
graphene's honeycomb lattice in such a way that the $x$ axis
corresponds to a zigzag direction. In this case, the gauge field
\emph{A} acting on charge carriers in graphene can be written
as\cite{SA02b,M07}
\begin{align}
 A_x &= \pm  c
\frac{\beta}{a} \left( u_{xx} - u_{yy} \right) \nonumber \\ A_y &=
\mp 2 c \frac{\beta}{a} u_{xy}    \label{gauge}
\end{align}
where $\beta = -\partial \log ( t ) / \partial \log ( a )$, where
$t \approx 3$eV is the electron hopping between $p_z$ orbitals
located at nearest neighbor atoms, $a \approx 1.4$\AA \, is the
distance between them, $c$ is a numerical constant that depends on
the details of atomic displacements within the lattice unit cell,
and $u_{ij}$ is the strain tensor. The two signs correspond to the
two valleys, $K$ and $K'$ in the Brillouin zone of graphene.

In two dimensional elasticity problems, it is convenient to study
the stress tensor, $\sigma_{ij} = \partial {\cal F} / \partial
u_{ij}$, where ${\cal F}$ is the elastic energy\cite{LL59}. The
gauge field can be  written in terms of the stress tensor as
\begin{align}
 A_x &= \pm  c \frac{\beta}{2 a \mu} \left( \sigma_{xx} - \sigma_{yy}
\right)   \nonumber \\ A_y &= \mp c \frac{\beta}{\mu a} \sigma_{xy}
\end{align}
where $\mu$ is a Lam\'e coefficient. Furthermore, possible stress
distributions that describe two dimensional elastic systems in
equilibrium can be written in terms of complex variables $z = x + i
y$ and $\bar{z} = x - i y$ as\cite{LL59}
\begin{align}
\sigma_{xx} &= \frac{\partial^2 f ( z , \bar{z} )}{\partial y^2}
\nonumber \\
\sigma_{yy} &= \frac{\partial^2 f ( z , \bar{z} )}{\partial x^2}
\nonumber \\
\sigma_{xy} &= - \frac{\partial^2 f ( z , \bar{z} )}{\partial x
\partial y}
\end{align}
Here $f ( z , \bar{z} )$ is either the real or the imaginary part
of a function
\begin{equation}
{\cal F} ( z , \bar{z} ) = {\cal F}_1 ( z ) + \bar{z} {\cal F}_2 ( z
)
\end{equation}
where ${\cal F}_1 ( z )$ and ${\cal F}_2 ( z )$ are analytic
functions. For the case of pure shear deformations considered in
Ref.\onlinecite{GKG09}  ${\cal F}_2 = 0$ but here we will not
restrict ourselves by this limitation. Since both stress and
\emph{A} are given by the second derivatives of ${\cal F}$,
whereas $B_{S}$ is given by the first derivatives of \emph{A}, a
uniform $B_{S}$ necessitates ${\cal F}$ to have a cubic dependence
on the coordinates. Such a function must have the following
structure:
\begin{equation}
{\cal F} ( x , y ) = c_1 ( x + i y )^3 + c_2 ( x - i y ) ( x + i y
)^2
\end{equation}
where $c_1$ and $c_2$ are arbitrary constants. Separating the real
and imaginary part of equation (6), we find four possible functions
that  result in uniform $B_{S}$
\begin{equation}
f( x , y ) \propto \left\{ \begin{array}{c}  x^3 - 3 x y^2 \\  x^3 +
x y^2
\\  3 x^2 y - y^3 \\ x^2 y + y^3 \end{array} \right.
\label{solutions}
\end{equation}
The second pair of the solutions is equivalent to the first one by
swapping the axes. For the lattice orientation used in
Eq.~(\ref{gauge}), the first pair leads to a gauge field such that
$A_x \propto x$ and $A_y \propto y$ and, accordingly, $B_{S}$ is
zero. Hence, the stress distributions that give rise to a uniform
pseudomagnetic field can be expressed in terms of a superposition of
the functions in lines 3 and 4. The strain configuration found in
ref.\onlinecite{GKG09} involves only the 3rd function $f(x,y)
\propto 3 x^2 y - y^3$, which leads to a unique solution for the
shape of graphene flake where such distribution of stresses can be
created by normal forces only. Unfortunately, this solution is not
easy to realize experimentally. The use of both 3rd and 4th
functions offers further possibilities.

\begin{figure}
\includegraphics[width=6cm]{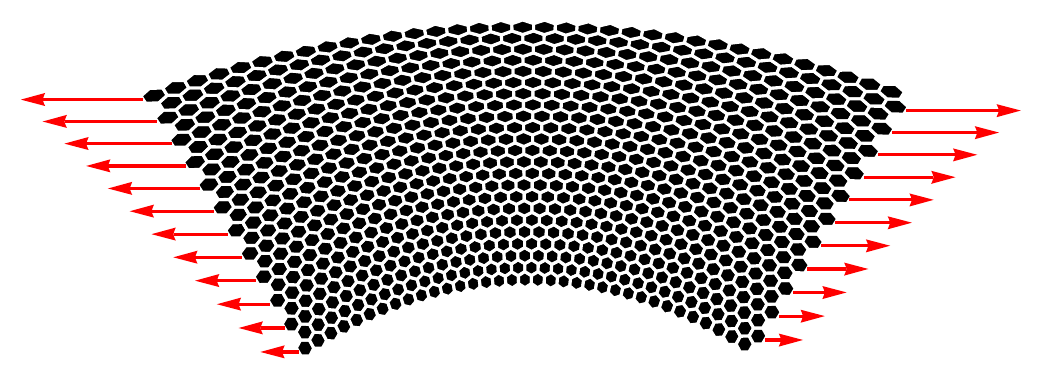}
\caption[fig]{(Color online). Stretching geometry leading  to a
uniform pseudomagnetic field inside a rectangular graphene sample.
Normal forces are applied at two opposite boundaries and their
magnitude is indicated by the length of the plotted arrows.}
 \label{bent_layer}
\end{figure}

In the following, we consider the deformations required to create
a uniform $B_{S}$ inside a rectangular graphene crystal, of width
$W$ and length $L$, with normal forces applied at the left and
right boundaries as sketched in Figure~\ref{bent_layer}. The
non-deformed crystal fills the region $-L/2 \le x \le L/2 , -W/2
\le y \le W/2$. Let us write the forces at the right and left
boundaries $x = \pm L/2$ as
\begin{align}
F_x^{R,L} &=  f_0^{R,L} + f_1^{R,L} y \nonumber \\
F_y^{R,D} &= 0
\end{align}
The condition of zero total force and zero total torque requires
$f_0^{R} = - f_0^L = f_0$ and $f_1^R = - f_1^L = f_1$ where  $f_0$
and $f_1$ are constants. The absence of forces at the upper and
lower edges implies that $\sigma_{yy} = \sigma_{xy} = 0$ there. At
the right and left edges, we have
\begin{align}
\sigma_{xx} &= \sigma_0 \left( y + \frac{f_0}{f_1} \right) \nonumber \\
\sigma_{xy} &= 0 \label{boundary}
\end{align}
where $\sigma_0$ is a constant that depends on applied forces. A
stress distribution within the crystal, which is compatible with
these boundary conditions, is generated by a function $f ( y ) = C [
y^3 / 3 + ( f_0 / f_1 ) ( y^2 / 2 ) ]$. This function can be
considered as a superposition of solutions 3 and 4  of
Eq.~(\ref{solutions}), which leads to a uniform $B_{S}$, and a
constant term that describes a uniaxial strain and does not give
rise to an additional pseudomagnetic field. The latter term ensures
that the lattice is stretched everywhere, there is no possibility
for out of plane deformations. Inside the rectangular crystal,
$\sigma_{xx}$ depends only on $y$ in the manner described by
eq.~(\ref{boundary}), and $\sigma_{yy} = \sigma_{xy} = 0$.  From
this stress distribution, we find the lattice distortions
\begin{align}
u_x &= u_0 \left( 2 x y + \frac{f_0}{f_1} x \right) \nonumber
\\
u_y &= u_0 \left[ - x^2 - \frac{\lambda}{\lambda + 4 \mu} \left( y^2
+ \frac{f_0}{f_1} y \right) \right] \label{desp}
\end{align}
where $u_0$ is a constant that defines the maximum stress. These
displacements lead to the curved shape shown in
Fig.~\ref{bent_layer}, which was drawn using the
reported\cite{ZKF09} Lam\'e coefficients of graphene, $\lambda
\approx 3.3$eV \AA$^{-2}$ and $\mu \approx 9.4$eV \AA$^{-2}$. The
maximum strain occurs at the top and bottom boundaries and can be
estimated as $\bar{u}_{max} \approx u_0 ( W + f_0 / f_1 )$. The
pseudomagnetic field inside the graphene crystal is given by
\begin{align}
B_{S} &= c \beta \frac{2 \Phi_0 u_0}{a} = c \beta \frac{\Phi_0
\bar{u}}{a W}
\label{estimate}
\end{align}
where $\Phi_0$ is the flux quantum. The effective magnetic  length
is $\ell_B = \sqrt{(a W)/(\beta \bar{u})}$. This field has the
same dependence on the crystal dimensions and the maximum strain
as in the examples discussed in Ref.\onlinecite{GKG09}. For $W
\approx 0.1$ micron and $\bar{u} \approx 10 \%$ the generated
effective field is of the order of 20T.

Experimentally, it may be difficult to create the precise  stress
distribution prescribed by Eq.~(\ref{boundary}). However, one can
see that the required shape of the strained graphene crystal in
Fig.~2 resembles an arc of a circle. To this end, we consider next
the geometry in which a rectangular graphene crystal is bent into a
circular arc, as sketched in Fig.~1 and shown in more detail in
Fig.~\ref{bent_substrate_circ}a. Note that this geometry is in fact
standard for experimental studies of strain (see, e.g.,
Ref.\onlinecite{Metal09,Hetal09,Tetal09b}) with the only difference
that the bending should be applied in plane rather than out of plane
of a graphene sheet.

If the radius of the inner circle defining the lower edge  in
Fig.~3a is $R$, the shape of the deformed rectangle is given by
\begin{align}
u_x ( x , y ) &= ( R + y ) \sin \left[ \frac{2 x}{L} \arcsin \left(
\frac{L}{2 R} \right) \right] - x \nonumber \\
u_y ( x , y ) &= ( R + y ) \cos \left[ \frac{2 x}{L} \arcsin \left(
\frac{L}{2 R} \right) \right] - R - y \label{circ}
\end{align}
The undistorted rectangular shape is recovered for $L/R
\rightarrow 0$.  The displacements in Eq.~(\ref{circ}) can be
expanded in powers of $R^{-1}$, and the leading terms are
\begin{align}
u_x ( x , y ) &= \frac{xy}{R} \nonumber \\
u_y ( x , y ) &= - \frac{x^2}{2 R} \label{expansion}
\end{align}
These displacements do not exceed ${\rm max} ( L , W )^2 / R$. The
next terms lead to corrections bound by ${\rm max} ( L , W )^3 /
R^2$. The distortions in Eq.~(\ref{expansion}) coincide with those
in eq.~(\ref{desp}) in the limit of vanishing Poisson ratio
$\lambda / ( \lambda + 2 \mu ) \rightarrow 0$  and lead to a
uniform $B_{S}$ inside the sample. The maximum strain is $L / 2 R
$. For an arbitrary value of $L/R$, the pseudomagnetic field is
given by
\begin{align}
B_{S} ( x , y ) &= - 4 c \frac{\beta \Phi_0}{a L } \arcsin \left(
\frac{L}{2 R} \right) \cos \left[ \frac{2 x}{L} \arcsin \left(
\frac{L}{2 R} \right) \right] \times \nonumber \\ &\times \left[ 1 -
\frac{R+y}{L} \arcsin \left( \frac{L}{2 R} \right)
\right]\label{precise}
\end{align}

\begin{figure}
\includegraphics[width=3.5cm,angle=90]{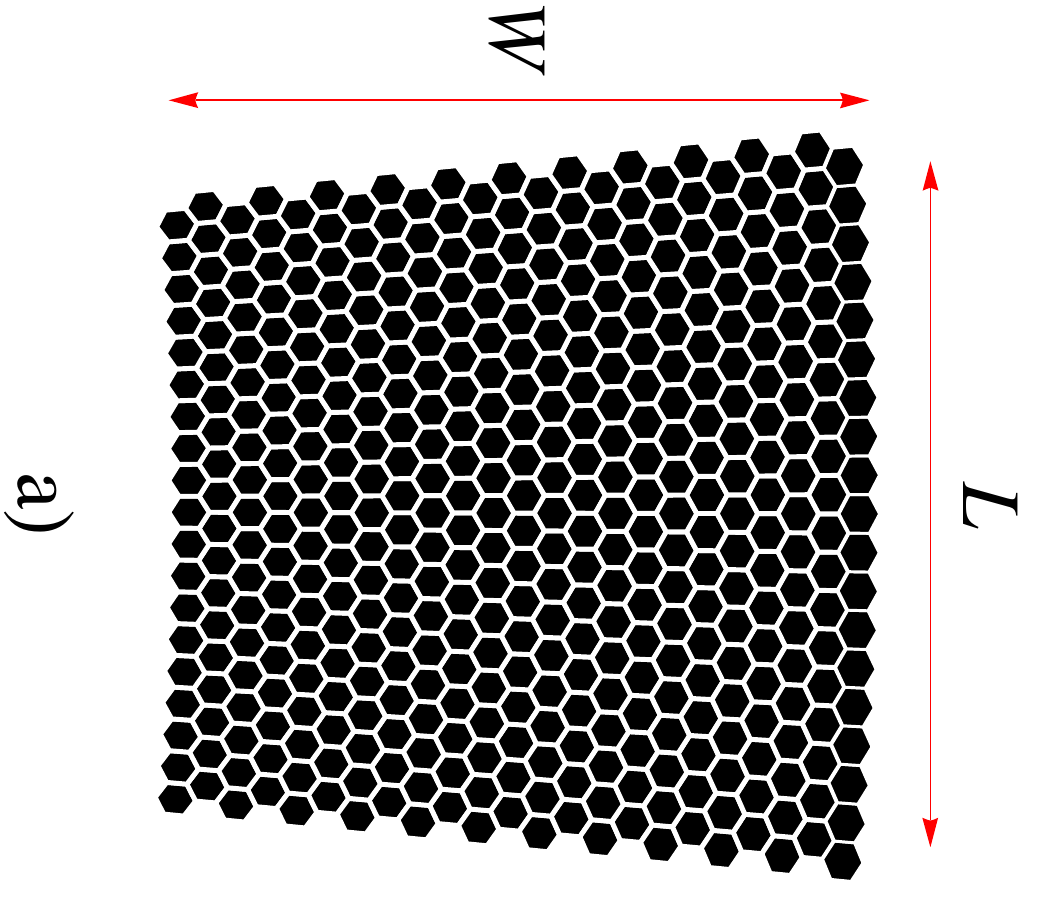}
\includegraphics[width=3.5cm]{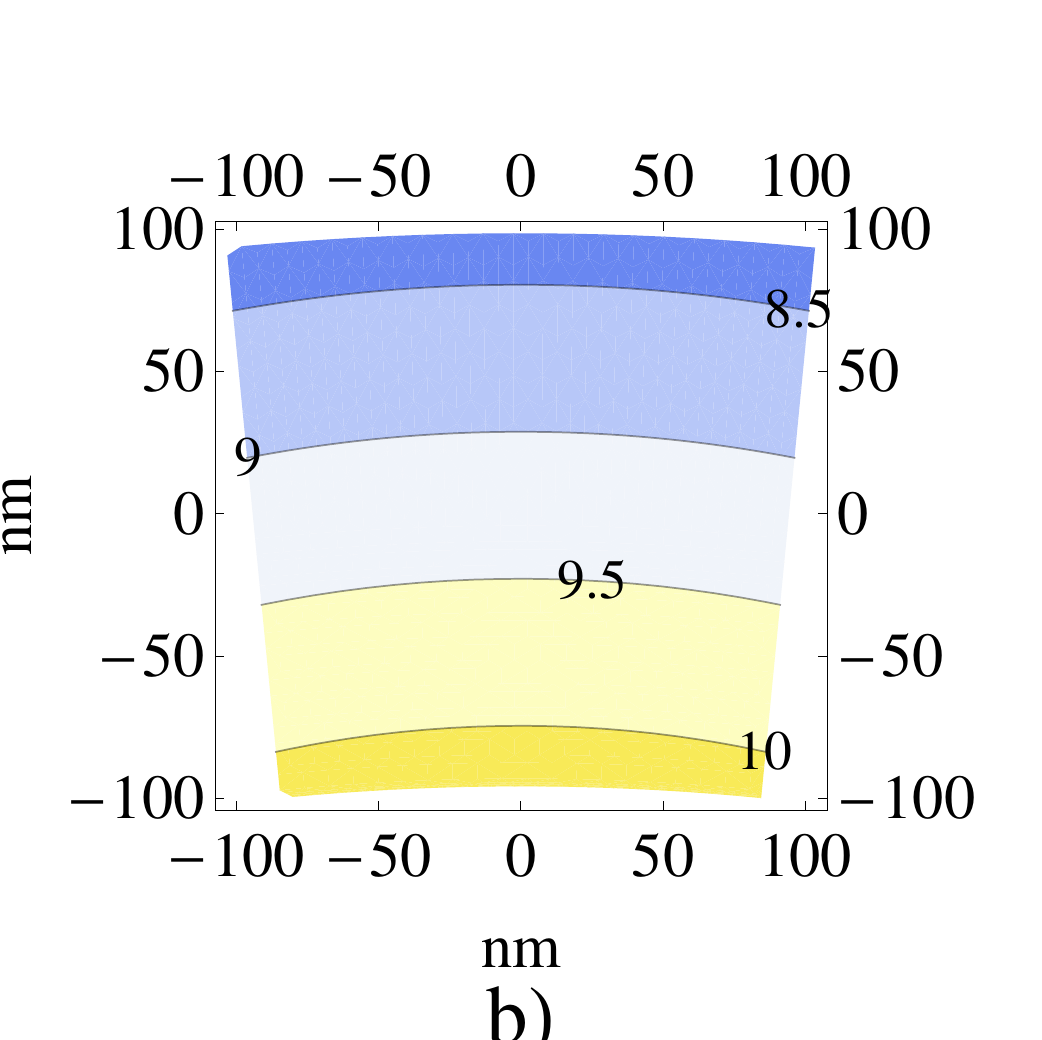}
\caption[fig]{(Color online). (\textbf{a}) - Rectangular graphene
sample deformed into an arc. The radii of the lower and upper edges
are $R$ and $R+W$, respectively. The plot is for $R=5 \times L$.
(\textbf{b}) - Effective magnetic field, in Teslas, for the same
deformed ribbon. Dimensions are $W = 200$nm, $L = 192$nm, $R = 5
\times L = 960$ nm.  The maximum strain is 10$\%$.}
\label{bent_substrate_circ}
\end{figure}

For $L/R \rightarrow 0$, the field reduces to $B_{S}  \approx -c (
\beta \Phi_0 )  / (a R )$, in agreement with Eq.~(\ref{estimate})
and the estimates given in Ref.\onlinecite{GKG09}. The relative
corrections to the constant value of $B_{S}$ are of the order $L /
2 R $, that is, the maximum strain. An example of the field
distribution described by Eq.~(\ref{precise}) is plotted in
Fig.~\ref{bent_substrate_circ}b.

The found strain is not purely shear but it also contains a
dilatation. The latter gives rise to an effective scalar
potential\cite{SA02b}, in addition to the discussed pseudomagnetic
field. Below, we show that due to screening, the extra potential
does not radically affect Landau quantization.

Following Ref.\onlinecite{SA02b}, and using eqs.~\ref{desp}, this
potential is:
\begin{equation}
V ( x , y ) = V_0 \left( \partial_x u_x + \partial_y u_y \right) =
V_0 \left[ 2 u_0 y + \frac{4 \mu f_0}{f_1 ( \lambda + 4 \mu )}
\right] \label{scalar}
\end{equation}
where $V_0 \approx 3$eV, estimated from the linear rise in the work
function of graphene under compression between 0 and 10\%
strain\cite{CJS09} (note that in Ref.\onlinecite{SA02b} a much
larger value of 16 eV is quoted, from old experimental data on
transport properties of graphite). The constant contribution gives a
rigid shift to all the levels while the non uniform term is
equivalent to an effective electric field along the $y$ direction.
Unlike the strain induced gauge field, this potential will induce a
charge redistribution and will be screened by the carriers in
graphene.

We consider first the screening expected if we assume that the
flake is a perfect metal, neglecting corrections due to quantum
properties of the electron gas. The induced charge density,
$\delta \rho ( \vec{r} )$ can be thus obtained from the condition
\begin{equation}
e^2 \int d^2 \vec{r}' \frac{\delta \rho ( \vec{r}' )}{| \vec{r} -
\vec{r}' |} = \frac{2 V_0 \bar{u}}{W} \delta \rho ( \vec{r} ) y
\label{elec}
\end{equation}
This equation can be rescaled by the substitution $\delta \bar{\rho}
( x / L , y / W ) = ( W + L )/(2 L ) \times ( 2 V_0 \bar{u} ) / (
e^2 W ) \times \delta \rho ( x , y )$.
%\begin{align}
%\bar{x} &= \frac{x}{L} \nonumber \\
%\bar{y} &= \frac{y}{W} \nonumber \\
%\delta \bar{\rho} \left( \frac{x}{L} , \frac{y}{W} \right) &=
%\frac{W + L}{2 L} \frac{2 V_0 \bar{u} }{e^2 W} \delta \rho ( x , y )
%\label{scaling}
%\end{align}
where the function $\delta \bar{\rho} ( \bar{x} , \bar{y} )$ depends
only on the aspect ratio, $W/L$.
\begin{figure}
\includegraphics[width=4cm]{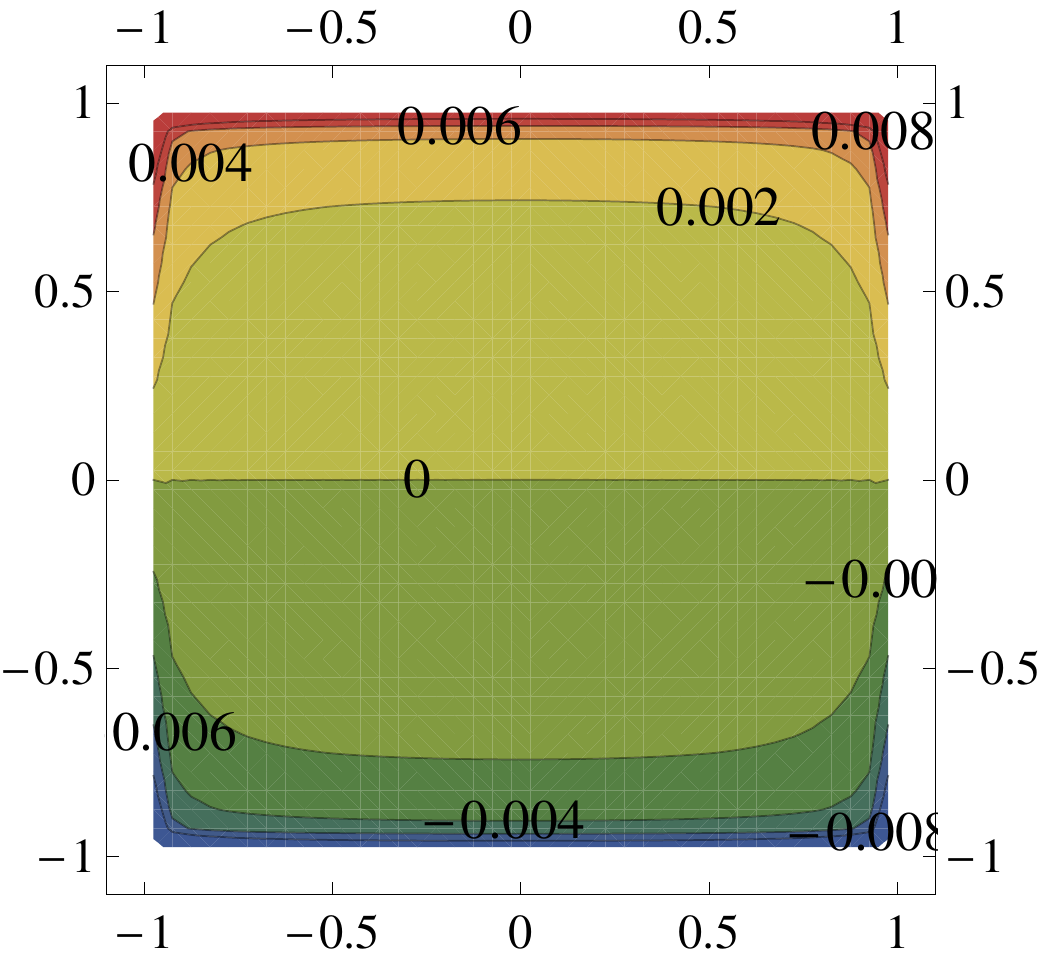}
\caption{(Color online). Variation in density, $\delta \bar{\rho}
( \bar{x} , \bar{y} )$, in dimensionless units (see text) due to
the screened scalar potential induced by strains. The aspect ratio
is $L/W=1$.} \label{fig_scalar}
\end{figure}

The density of carriers in a given Landau level due to the effective
field given in Eq.~\ref{estimate}, setting $c=1$, is $\rho_{LL} = 2
( a W ) / ( \beta \bar{u} )$ so that $\delta \rho / \rho_{LL} =
(W+L)/(2L) \times (4 V_0 a )/(\beta e^2) \times \delta \bar{\rho} (
x/L , y/W )$.
%\begin{equation}
%\frac{\delta \rho ( x , y )}{\rho_{LL}} = \frac{W+L}{2L} \frac{4 V_0
%a}{\beta e^2} \delta \bar{\rho} \left( \frac{x}{L} , \frac{y}{W}
%\right)
%\end{equation}
%The factor $( V_0 a ) / ( e^2 \beta )$ is of order unity, so that
Hence, the non uniform density induced by the effective field, in
units of the difference in densities between different quantum Hall
plateaus,  is approximately given by $\delta \bar{\rho}$ shown in
Fig.~\ref{fig_scalar}. %The results change slightly with the aspect
%ratio of the flake.

To provide an ideal metallic screening, the chemical potential
should coincide at each point with the positions of one of the
Landau levels. Hence, the quantum energy of the carriers is larger
than in a uniform electron distribution, when the Fermi energy lies
in a pseudogap between the Landau levels. Thus, in addition to the
classical screening energy, discussed above, we also must add the
change in energy due to the changes in occupancies of the electron
levels in the presence of the scalar potential. We first assume that
the induced electronic density is given by the solution of
Eq.~\ref{elec}, and that the electronic states are the Landau levels
induced by the effective field in Eq.~\ref{estimate}. Then, the two
contributions to the energy, for $W \sim L$, are of the order:
\begin{align}
E_{elec} &\sim \frac{( V_0 \bar{u} )^2 L}{e^2} \langle | \delta
\bar{\rho} | \rangle^2
\nonumber \\
E_{quantum} &\sim \frac{v_F \beta V_0 \bar{u}^2}{e^2} \langle |
\delta \bar{\rho} | \rangle \sqrt{\frac{L}{a}}
\end{align}
The calculations shown in Fig.~\ref{fig_scalar} suggest that
$\langle | \delta \bar{\rho} | \rangle = f \approx 10^{-2} -
10^{-1}$, and $v_F \sim e^2 \sim V_0 a$. Then, the scalar potential
is screened, and the process can be described by the classical model
outlined earlier, if $\sqrt{L/a} \gtrsim f^{-1}$, that is, $L
\gtrsim 10^2 - 10^3$nm. For smaller sizes, the rigidity of the
quantum levels induced by the gauge potential prevents any
rearrangement of the charge inside the flake. A detailed theory of
screening in this situation will be presented elsewhere. In either
case, the electronic states are well described by the effective
Landau levels induced by the field in Eq.~\ref{estimate}.

To create the required strain experimentally,  one can
think of depositing graphene ribbons onto a rectangular elastic
substrate and deform it in the manner prescribed by eq.~(9) or by
bending it into a circular arc (Fig.~1). Crystals rigidly
attached to the substrate can then be of arbitrary shape, as the
strain distribution in the substrate would project onto graphene and
give rise to a (nearly) uniform $B_{S}$. Note that macroscopic substrates
would require the use of rubber-like materials capable of withstanding
very large strains, such that local deformations projected on a submicron
graphene crystals could still reach ~$\approx 10\%$.

$Acknowledgements$ - FG acknowledges support from MICINN (Spain)
through grants FIS2008-00124 and CONSOLIDER CSD2007-00010, and by
the Comunidad de Madrid, through CITECNOMIK. MIK acknowledges
support from FOM (the Netherlands). This work was also supported by
EPSRC (UK), ONR, AFOSR, and the Royal Society. We are thankful to
Y.-W. Son for useful insights concerning~\cite{CJS09} and related
work.

\bibliography{bib_suspended_2}
\end{document}